\documentclass[a4paper, amsfonts, amssymb, amsmath, reprint, showkeys, nofootinbib, twoside, superscriptaddress]{revtex4-1}
\usepackage[english]{babel}
\usepackage[utf8]{inputenc}
\usepackage[colorinlistoftodos, color=green!40, prependcaption]{todonotes}
\usepackage{amsthm}
\usepackage{mathtools}
\usepackage{physics}
\usepackage{xcolor}
\usepackage{graphicx}
\usepackage[left=23mm,right=13mm,top=35mm,columnsep=15pt]{geometry} 
\usepackage{adjustbox}
\usepackage{placeins}
\usepackage[T1]{fontenc}
\usepackage{lipsum}
\usepackage{csquotes}
\usepackage[pdftex, pdftitle={Article}, pdfauthor={Author}, hidelinks]{hyperref}
\usepackage{textalpha}
\DeclareUnicodeCharacter{2212}{-}

\begin{document}

\title{Near-Field Localization of the Boson Peak on Tantalum Films for Superconducting Quantum Devices}

\author{Xiao Guo}
    \altaffiliation{Both authors contributed equally to this work.}
    \affiliation{School of Information Technology and Electrical Engineering, The University of Queensland, Brisbane, QLD 4072, Australia}
\author{Zachary Degnan}
    \altaffiliation{Both authors contributed equally to this work.}
    \affiliation{School of Mathematics and Physics, The University of Queensland, Brisbane, QLD 4072, Australia}
    \affiliation{ARC Centre of Excellence for Engineered Quantum Systems, Brisbane, QLD 4072, Australia}
\author{Julian Steele}
    \affiliation{School of Mathematics and Physics, The University of Queensland, Brisbane, QLD 4072, Australia}
\author{Eduardo Solano}
    \affiliation{NCD-SWEET Beamline, ALBA Synchrotron Light Source, 08290, Cerdanyola del Vallès, Barcelona, Spain.}
\author{Bogdan C. Donose}
    \affiliation{School of Information Technology and Electrical Engineering, The University of Queensland, Brisbane, QLD 4072, Australia}
\author{Karl Bertling}
    \affiliation{School of Information Technology and Electrical Engineering, The University of Queensland, Brisbane, QLD 4072, Australia}
\author{Arkady Fedorov}
    \affiliation{School of Mathematics and Physics, The University of Queensland, Brisbane, QLD 4072, Australia}
    \affiliation{ARC Centre of Excellence for Engineered Quantum Systems, Brisbane, QLD 4072, Australia}
\author{Aleksandar D. Raki\'{c}}
    \email[email: ]{a.rakic@uq.edu.au}
    \affiliation{School of Information Technology and Electrical Engineering, The University of Queensland, Brisbane, QLD 4072, Australia}
\author{Peter Jacobson}
    \email[email: ]{p.jacobson@uq.edu.au}
    \affiliation{School of Mathematics and Physics, The University of Queensland, Brisbane, QLD 4072, Australia}
    \affiliation{ARC Centre of Excellence for Engineered Quantum Systems, Brisbane, QLD 4072, Australia}


\begin{abstract}
Superconducting circuits are among the most advanced quantum computing technologies, however their performance is currently limited by losses found in surface oxides and disordered materials. Here, we identify and spatially localize a near-field signature of loss centers on tantalum films using terahertz scattering-type scanning near-field optical microscopy (s-SNOM). Making use of terahertz nanospectroscopy, we observe a localized excess vibrational mode around 0.5~THz and identify this resonance as the boson peak, a signature of amorphous materials. Grazing-incidence wide-angle x-ray scattering (GIWAXS) shows that oxides on freshly solvent-cleaned samples are amorphous, whereas crystalline phases emerge after aging in air. By localizing defect centers at the nanoscale, our characterization techniques and results will inform the optimization of fabrication procedures for new low-loss superconducting circuits. 
\end{abstract}

\keywords{tantalum, scanning near-field optical microscopy, terahertz spectroscopy, grazing incidence wide angle x-ray scattering, superconducting quantum devices, amorphous materials}

\maketitle


The current central goal of superconducting quantum computing is to improve the fidelity of qubits enough to implement operations with minimal error correction~\cite{bravyi_future_2022}. Qubit fidelity, and in turn the progress of quantum computing, is currently limited by decoherence due to the dissipative coupling of qubit modes to electric dipoles in amorphous materials and defect states~\cite{you_stabilizing_2022, siddiqi_engineering_2021}. These loss channels concentrate at amorphous oxides present at device interfaces, such as the metal-air interface~\cite{lisenfeld_electric_2019}, and their removal significantly improves the performance of planar superconducting devices in the low-temperature and low-power regimes~\cite{earnest_substrate_2018}. It is therefore critical to devise methods to minimize oxidation and/or control the chemical reactivity of metal surfaces found in quantum devices~\cite{de_leon_materials_2021, de_graaf_chemical_2022}.

Tantalum has been identified as the leading material system for the fabrication of superconducting devices with state-of-the-art performance, showing notable improvements over aluminum or niobium-based devices~\cite{place_new_2021, wang_towards_2022}.
While initial results are promising, surface oxides remain the primary factor limiting device performance~\cite{wang_towards_2022, lozano_manufacturing_2022}.
Tantalum oxide (Ta$_2$O$_5$) has well-documented crystalline and amorphous phases, and both phases possess similar coordination shells below 4~{\AA}~\cite{sahu_theoretical_2004, bassiri_order_2015, shyam_measurement_2016, martinelli_deep_2021}.
%
%
%
The main structural difference is a reduced Ta coordination number in the amorphous phase ($a\textrm{-}\textrm{Ta}_2\textrm{O}_5$)~\cite{martinelli_deep_2021}.
%
%
For native amorphous oxide layers atop the Ta surfaces, it is unclear whether the structural trends observed for bulk oxide phases persist at the metal-air interface~\cite{sasikumar_evolutionary_2017}.
While the structural motifs at these interfaces are less explored, electronic structure measurements using (soft) x-ray photoelectron spectroscopy after near-ambient oxidation indicate pentoxide and suboxide phases are present, increasing the chemical complexity of this amorphous oxide~\cite{wang_situ_2010}.

Amorphous materials have well-documented anomalous thermodynamic behaviors at low temperatures such as deviations from the Debye model~\cite{zeller_thermal_1971, zhang_experimental_2017, binder_glassy_2011}.
This behavior was explained independently by Phillips and Anderson \textit{et al.}, who proposed a model of tunneling two-level systems (TLS; Figure 1a)~\cite{phillips_tunneling_1972, anderson_anomalous_1972}.
For superconducting devices and hardware (resonators, Josephson junctions etc.), TLSs are a ubiquitous source of noise and fidelity loss~\cite{muller_towards_2019}. However, removing TLS is not straightforward as TLS are heterogeneous with structural motifs ranging from single atoms to groups of atoms, which tunnel between configurations of differing energy.
Major successes of the TLS model include capturing the effect of strain on individual TLSs in active superconducting devices~\cite{lisenfeld_electric_2019, grabovskij_strain_2012}, and reproducing peaks in the heat capacity at low temperatures due to an excess vibrational density of states, i.e. the boson peak~\cite{binder_glassy_2011}.
The boson peak is a universal signature of amorphous states of matter --- ranging from disordered aggregates of colloidal nanoparticles~\cite{chen_low-frequency_2010} and organic materials~\cite{angell_formation_1995, hong_pressure_2008}, to metallic glasses~\cite{luo_memory_2016}.
Importantly, as TLS and the boson peak are ultimately lattice vibration phenomena, it can be detected spectroscopically and at elevated temperatures~\cite{binder_glassy_2011}.
Probes sensitive to the boson peak include inelastic neutron scattering~\cite{nemanich_low-frequency_1977, buchenau_low-frequency_1986, cortie_boson_2020}, (hyper) Raman spectroscopy~\cite{winterling_very-low-frequency_1975, malinovsky_nature_1986}, helium atom scattering~\cite{steurer_observation_2007, tomterud_observation_2022}, and terahertz (THz) time-domain spectroscopy~\cite{Shibata2015BonsonPeak, kabeya_boson_2016, mori_detection_2020, Naftaly2021BosonPeak}.
However, these methods are spatially averaged with responses restricted by (at best) the diffraction limit, precluding an understanding of the local spatial variation of TLS across microscopic surface regions.
Due to their direct influence on qubit performance, the local identification of amorphous oxide phases on tantalum-based quantum devices represents a scientifically challenging and technologically important \textit{terra incognita}.

Here, we study oxide phases at Ta surfaces in the near-field.
The coexistence of two surface phases is observed with both IR and THz near-field responses in scattering-type scanning near-field optical microscopy (s-SNOM)~\cite{ChenXZ2019Review, Cocker2021Review}.
THz nanospectroscopy shows that flat regions exhibit a low-frequency spectral signature characteristic of the boson peak, confirming it as an amorphous phase, whereas 1D nanoridges show characteristic signatures of dielectric relaxation.
The presence of thin amorphous and crystalline oxides is corroborated by grazing-incidence wide-angle x-ray scattering (GIWAXS), while topographic and chemical changes associated with etching treatments are tracked by atomic force microscope (AFM) and x-ray photoelectron spectroscopy (XPS).
THz nanospectroscopy after etching indicates that the amorphous oxide has been altered or removed.
Our experiments demonstrate that THz s-SNOM in tandem with GIWAXS can differentiate between crystalline and amorphous phases at the nanoscale, thereby informing the processing of materials for superconducting quantum computing.

\begin{figure*}
    \centering
    \includegraphics[width=\linewidth]{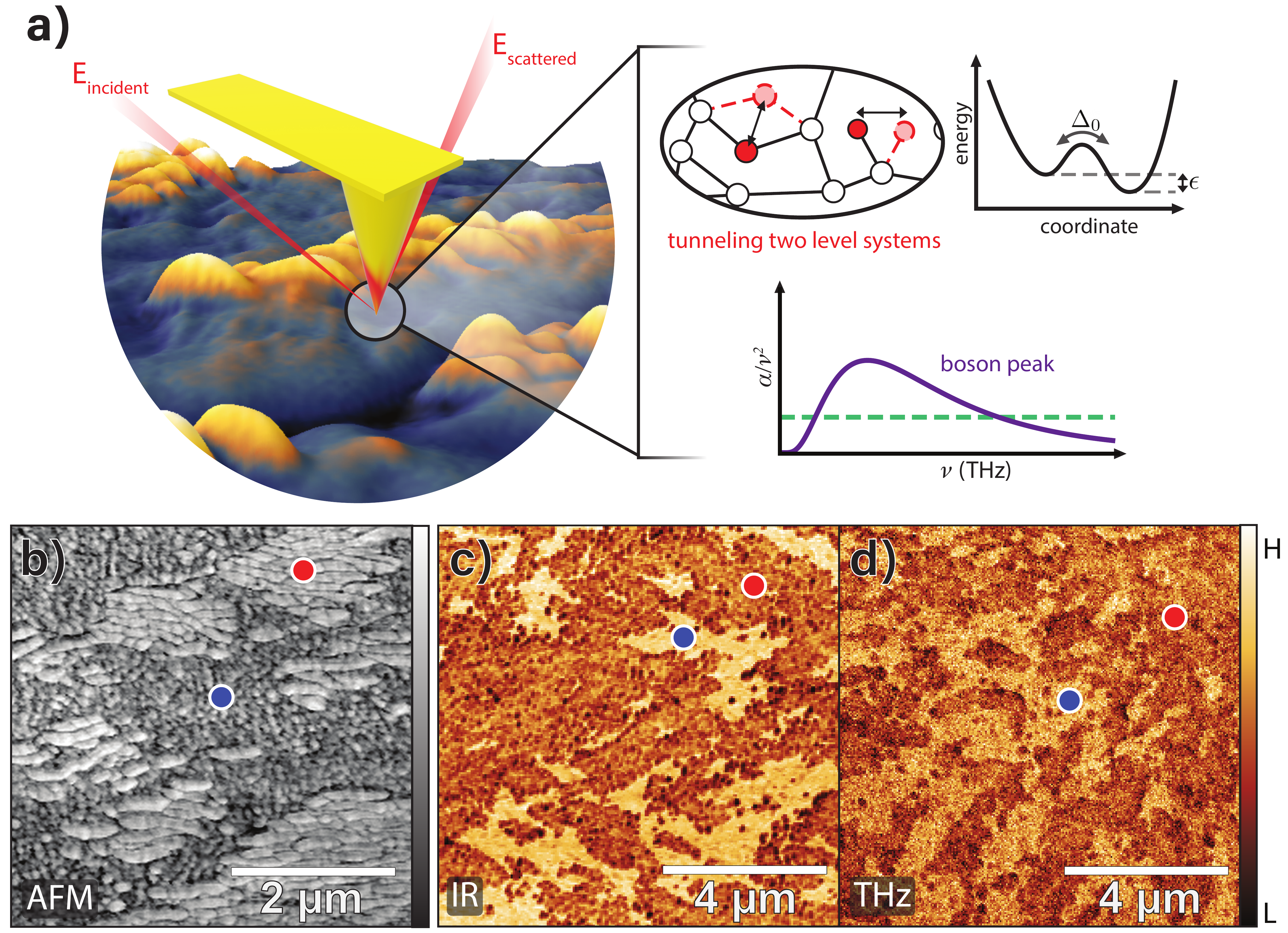}
    \caption{(a) Schematic operational principles of probing nanoscale light-matter interactions employing an s-SNOM (left): a broadband pulse (mid-IR/THz) is collimated onto a metallic probe tip periodically tapping on the sample surface with a nanofocus generated around the tip end to surpass the diffraction limit. The interrogated sample information is encoded into the tip-scattered field through the dipole-dipole interaction between the tip-sample pair. The demodulated scattering field at high-order harmonics ($n$-th) of the tip tapping frequency ($S_n$) highlights localized near-field responses at the nanoscale. The structure of an amorphous oxide (top middle) containing TLSs (red) in the form of individual/groups of atoms tunneling between two configurations. The TLSs are modeled by a double-well potential with an energy difference $E=\sqrt{\Delta_0^2 + \epsilon^2}$ between eigenstates, where $\Delta_0$ is the inter-well tunneling rate and $\epsilon$ is the asymmetry bias (top right)~\cite{phillips_two-level_1987}. Depiction of a boson peak with the characteristic log-normal distribution (solid purple) from THz nanospectroscopy, and a typical vibrational density of states curve for a Debye solid $\propto \nu^2$ (dashed green) (bottom right)~\cite{mori_detection_2020}. The surface of a solvent-cleaned Ta film interrogated by s-SNOM with (b) AFM phase, (c, d) the amplitude of third-harmonic s-SNOM scattering signals ($S_{3}$) from mid-IR and THz nanoimaging in the white-light mode.}
    \label{fig1}
\end{figure*}

We begin by isolating the topographic and near-field optical differences on Ta films using a combination of surface-sensitive probes.
Figure 1a schematically illustrates the s-SNOM measurement principle and the connection to TLS.
In s-SNOM, a metallic probe tip periodically taps the sample surface, which is simultaneously illuminated by an electromagnetic stimulus, e.g., THz radiation.
The probe tip is transiently polarized by the incident illumination and thereby forms a highly concentrated electric field --- a nanofocus --- near its apex.
With nanometer precision in the positioning of the nanofocus, and a probe tip radius around $60~\textrm{nm}$, THz s-SNOM is able to bypass the diffraction limit and resolve the nanoscale THz response as well as the sample topography in AFM.
Therefore, near-field imaging provides a spatial mapping of the spectrally-averaged optical response, while THz nanospectroscopy yields a frequency-dependent imprint of the material within the nanofocus~\cite{ChenXZ2019Review, Cocker2021Review}.
%
%

AFM of the Ta film reveals the surface consists of two distinct regions, an elevated striped region (A), and a low-lying region (B) (Figure 1b).
%
%
In near-field images (Figures 1c,d), two regions of strong optical scattering contrast with sharp boundaries are observed using both mid-IR and THz excitations.
%
%
%
The strongest scattered amplitude coincides with region B, while the stripes of region A show less prominent scattering indicating a difference in the dielectric response and implying a difference between the regions.
%

\begin{figure}
    \centering
    \includegraphics[width=\linewidth]{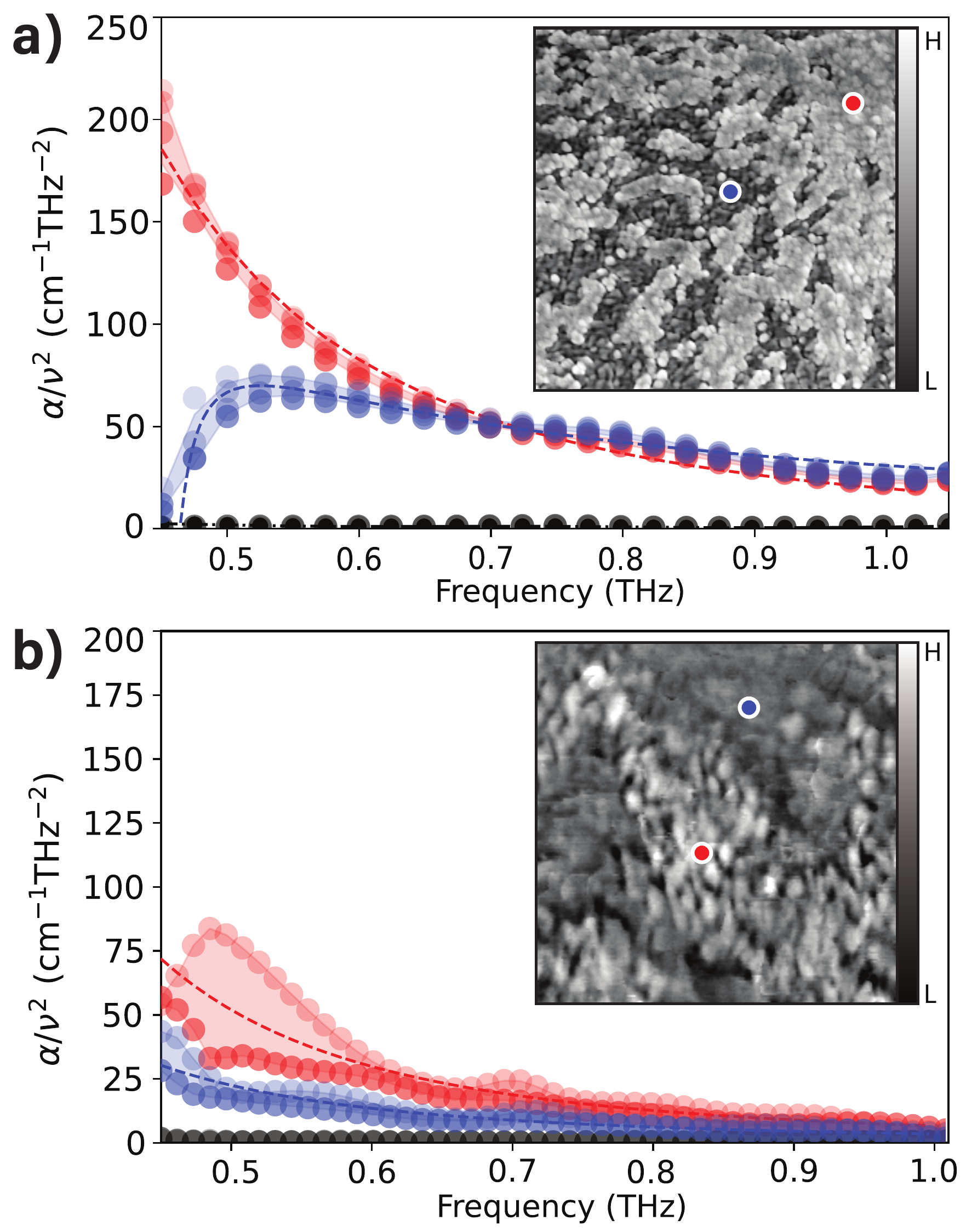}
    \caption{THz nanospectroscopy on solvent-cleaned and RIE-treated Ta surface. (a, b) Normalized absorption coefficient, $\alpha(\nu)/\nu^2$, from calibrated THz nanospectroscopy with multiple high-order harmonic signals ($S_n$, $n\geq 2$): The localized THz spectral absorption of region B (blue) follows a log-normal distribution as a characteristic boson peak feature for the unetched sample~(a) and turns to obey the power law after RIE surface etching~(b).  Localized THz spectral absorption in region A (red) follows a power law (red dashed lines, guided for eyes) towards decreasing THz frequencies. The spectral absorption feature for high-resistivity (non-metallic) silicon (black) is plotted for comparison. AFM phase images of the solvent-cleaned Ta surface (a), and after reactive ion etching (RIE) (b) are inset.}
    \label{fig2}
\end{figure}

To elucidate the origins of the difference between regions, THz nanospectroscopy was performed on both solvent-cleaned and reactive ion etched (RIE) Ta samples (Figure 2).
To suppress background noise in s-SNOM scattering spectra, higher-harmonic signals ($n \geq 2$) are used for s-SNOM vector calibration to quantitatively retrieve absorption coefficients, $\alpha(\nu)$, for region A and B (Figure S4)~\cite{guo_optical_2021}.
Normalized absorption coefficients ($\alpha(\nu)/\nu^2$) are obtained and compared with the Debye model, which appears as a horizontal line (Figure 1a, green dashed line).
%
For the solvent-cleaned sample, we observe a diverging response for regions A and B below $0.65$ THz and these trends are consistent for higher-order harmonic signals ($S_{3}$ – $S_{5}$) in both regions indicating exemplary signal to noise characteristics.
The relative permittivity ($\epsilon_r$) of Ta oxide ($>30$) is substantially greater than other common dielectrics (e.g. SiO$_2 \sim 5$) over this range~\cite{lee_measurement_2003}, facilitating the strong tip-scattered THz signals in s-SNOM.

The absorption in region A steeply increases with decreasing frequency, roughly following a power law dependence, indicating a reorientation of dipoles within a highly polarizable dielectric. 
In contrast, the absorption rapidly decreases in region B following a log-normal distribution at low frequency which is a characteristic boson peak signature~\cite{mori_detection_2020}.
%
%
%
For Ta samples processed by RIE (Figure 2b), the results are markedly different with an apparent overall reduction in the normalized absorption coefficient, attributed to the removal of dielectric material and a more metallic surface, as supported by XPS measurements and discussed below.     
%
%
Critically, this processing removes the spectroscopic signature due to the boson peak and the post-RIE response of both regions can be ascribed to dipole reorientations.

While there is robust debate on the precise cause and interpretation of the boson peak, it is understood as a universal indicator of materials with glassy or amorphous structures~\cite{grigera_phonon_2003, lubchenko_origin_2003, schirmacher_boson_2013, schirmacher_harmonic_1998, chumakov_equivalence_2011, shintani_universal_2008, hu_origin_2022}.
From the linear response theory for disordered materials, the absorption coefficient is proportional to the vibrational density of states (vDOS: $g(\nu)$) and the boson peak is usually characterized by $g(\nu)/\nu^2$~\cite{nemanich_low-frequency_1977, malinovsky_nature_1986}.
Recently, the boson peak has been observed and characterized using the normalized absorption coefficients $\alpha(\nu)/\nu^2$ from THz time-domain spectroscopic measurements on amorphous or glass-like materials~\cite{Shibata2015BonsonPeak, kabeya_boson_2016, mori_detection_2020, Naftaly2021BosonPeak}.
As Ta$_2$O$_5$ has well-documented amorphous phases, the spectroscopic signature at $0.5$ THz in region B fits with our understanding of the universal nature of the boson peak.
Therefore, our THz nanospectroscopy results indicate that $a$-TaO$_x$ exists at discrete locations on the surface of the film. 

\begin{figure*}
    \centering
    \includegraphics[width=\linewidth]{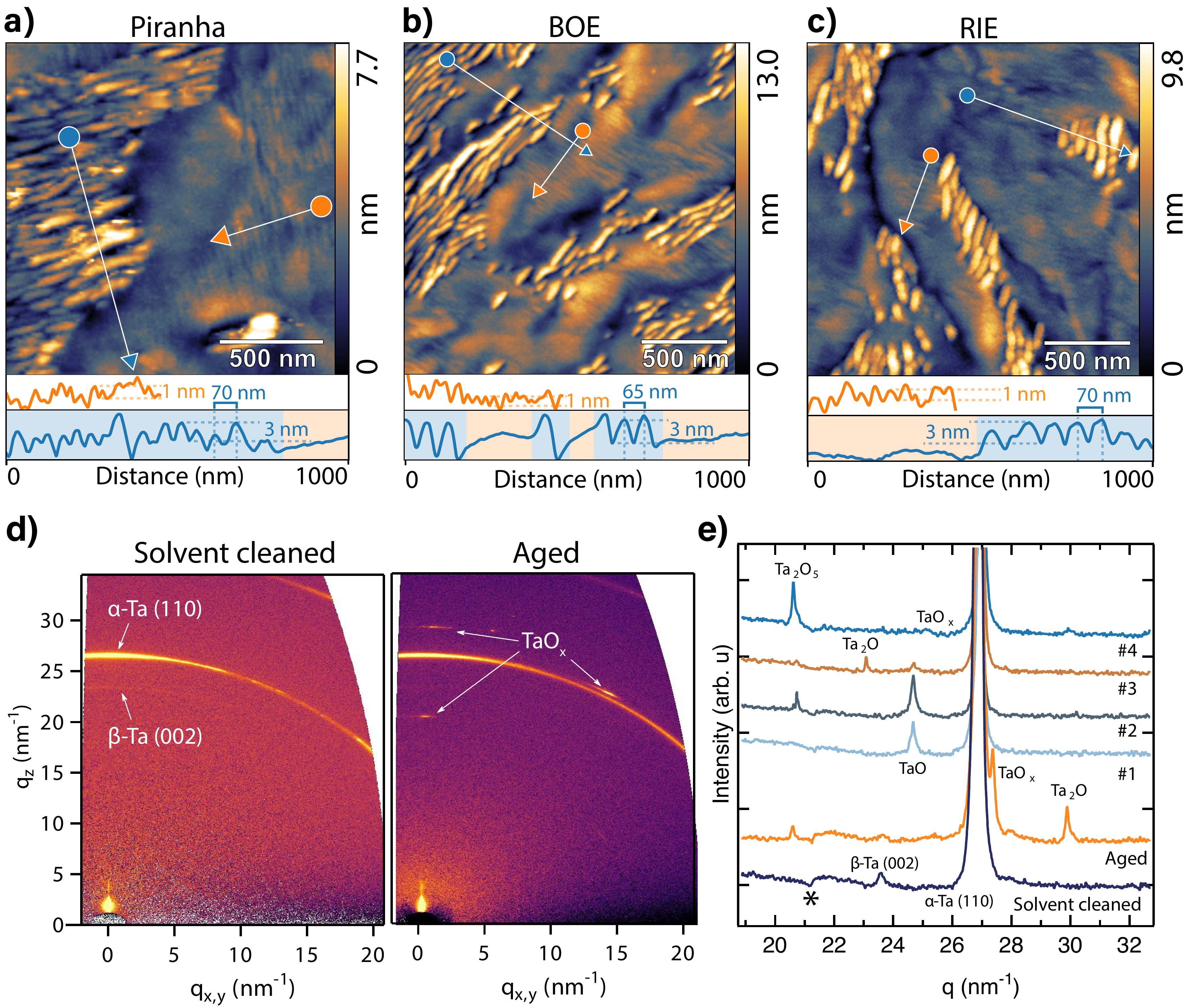}
    \caption{AFM and GIWAXS. AFM height images of Ta films after (a) piranha treatment, (b) buffered oxide etching (BOE), and (c) RIE. Line profiles are displayed below each AFM image, with the direction of each scan shown by the corresponding arrow in the image above. The blue scans traverse the 1D stripes of region A, and the orange scans move across the perpendicular stripes of region B. (d) 2D GIWAXS patterns recorded from a solvent-cleaned Ta film and an aged film exhibiting surface oxidation, due to exposure to ambient oxygen. (e) Integrated scattering signals ($q_{xyz}$) corresponding to the frames shown in (d), along with four additional examples ($\#$1-4) of varied TaO$_x$-related surface signals. These patterns have been expanded and rescaled known to contain TaO$_x$ peaks, with complete scattering patterns contained in Figure S7 for completeness. The asterisk ($\ast$) indicates an integration artifact, introducing a slight dip in the background.}
    \label{fig3}
\end{figure*}

As RIE processing removes the boson peak signature but retains the spatially varying near-field optical response, further surface treatments common to superconducting devices were evaluated.
High-aspect ratio AFM tips, in contrast to the larger radius tips used for THz and IR measurements, were employed to investigate the surface structure after piranha, BOE, and RIE (Figure 3).
On the piranha-cleaned sample, region A sits $\sim 3$~nm higher than region B (versus $\sim 1$~nm on solvent-cleaned), indicating that piranha cleaning removes material from region B.
The lateral spacing between parallel stripes in region A is $\sim 70$~nm and changing etching method (BOE, RIE) does not affect the structure.
Additionally, comparable surface coverages of region A (BOE, 24$\%$; piranha, 35$\%$; RIE, 28$\%$) are observed across treatments.

While region A is unchanged across etched and solvent-cleaned samples, we do observe topographic changes in region B.
Starting with the piranha-cleaned sample, region B is less rough than the solvent-cleaned sample and develops a weaker stripe pattern ($\sim 1$~nm corrugation) in place of the disordered structures observed in region B of the solvent-cleaned sample (Figure 3a, line profile).
Using the stripes of region A as a point of reference, these weaker stripes run nearly perpendicular to the 1D stripes.
We observe the same structure on BOE and RIE-treated samples, with stripes of similar corrugation and periodicity appearing in region B.
As all surface treatments lead to two surface regions with consistent topography – including RIE which removes material – the 1D stripes in region A represent structures formed during film nucleation. 

The prominent 1D stripes are consistent with growth modes observed in other bcc metals such as W where 1D ‘nanoridges’ are the result of anisotropic diffusion of adatoms at bcc (110) surfaces during film growth~\cite{singh_nanoridge_2003}.
This mass transport mechanism is corroborated by domain intersections, where the 1D stripes meet at angles of $\sim 110$ degrees, as expected for preferential diffusion on low index bcc surfaces (Figure S5).
However, mass transport alone cannot explain the persistent near-field contrast.
 Surface defects such as step edges or kinks are well-known reactive sites due to the reduced surface coordination and presence of adsorption sites not present on flat surfaces~\cite{ertl_reactions_2008}.
Highly stepped or microfaceted surfaces are often regions with enhanced chemical reactivity and more prone to oxidation through dissociation of molecular oxygen or water. 
Therefore, we attribute the strong near-field contrast at the surface of our films to the preferential oxidation of these 1D nanostructures.

Due to a variety of possible oxidation states (1+ to 5+), Ta surfaces exhibit complicated structural evolutions when exposed to ambient oxygen, kinetically transforming from a pure metal into a variety of intermediate metastable suboxides (TaO$_x$) before ultimately forming the thermodynamically stable pentoxide (Ta$_2$O$_5$).
The detection and identification of (sub)oxide phases subsequently require a phase-sensitive probe.
To elucidate the oxidation profile of our Ta film, we employ surface-sensitive GIWAXS (Figure S6) and scan through relatively shallow incident angles ($\alpha_i < 0.3^{\circ}$) to formulate a surface-subsurface structural model.
Conveniently, given that TaO$_x$ is relatively less dense than Ta, even a very thin oxide layer ($<5$~nm) can be studied due to the amplifying effects of surface refraction.

Figure 3d compares representative 2D GIWAXS images recorded from the solvent-cleaned and aged solvent-cleaned Ta films, with their corresponding integrated profiles contained in Figure 3e. 
Examining the full 1D profile (Figure S7), the fresh solvent-cleaned Ta material is shown to contain only a minor portion of metastable $\beta$-Ta that resides near the surface of a predominantly $\alpha$-Ta film (Figure S8). The Ta film is highly oriented with respect to the planar surface (Figure S9), a feature that is consistent across the entire film. Conversely, following 2 weeks of exposure to air, several new Bragg peaks are introduced within the scattering range between $q=18-33~\textrm{nm}^{-1}$~(Figure 3e) and we assess its development by comparing it to thermally driven oxidation~\cite{korshunov_oxidation_2020}. Here, the relatively prominent peak residing at $21.5$~nm$^{-1}$ is consistent with diffraction from the (110)/(200) scattering planes of the orthorhombic $\beta$-Ta$_2$O$_5$ structure, while splitting of the $\alpha$-Ta (110) peak indicates partial oxidation of Ta metal (TaO$_x$). An angle of incidence scan further demonstrates these signals arise from the uppermost surface of the film (Figure S8), and lateral sampling ($\sim 2$~mm) of different areas captures a variety of metastable surface oxides (Figure 2e). The thermodynamically preferred $\beta$-Ta$_2$O$_5$ phase is restricted to specific regions in the 2D GIWAXS pattern, indicating this phase grows in a highly faceted manner(Figure 3d).

\begin{figure*}
    \centering
    \includegraphics[width=\linewidth]{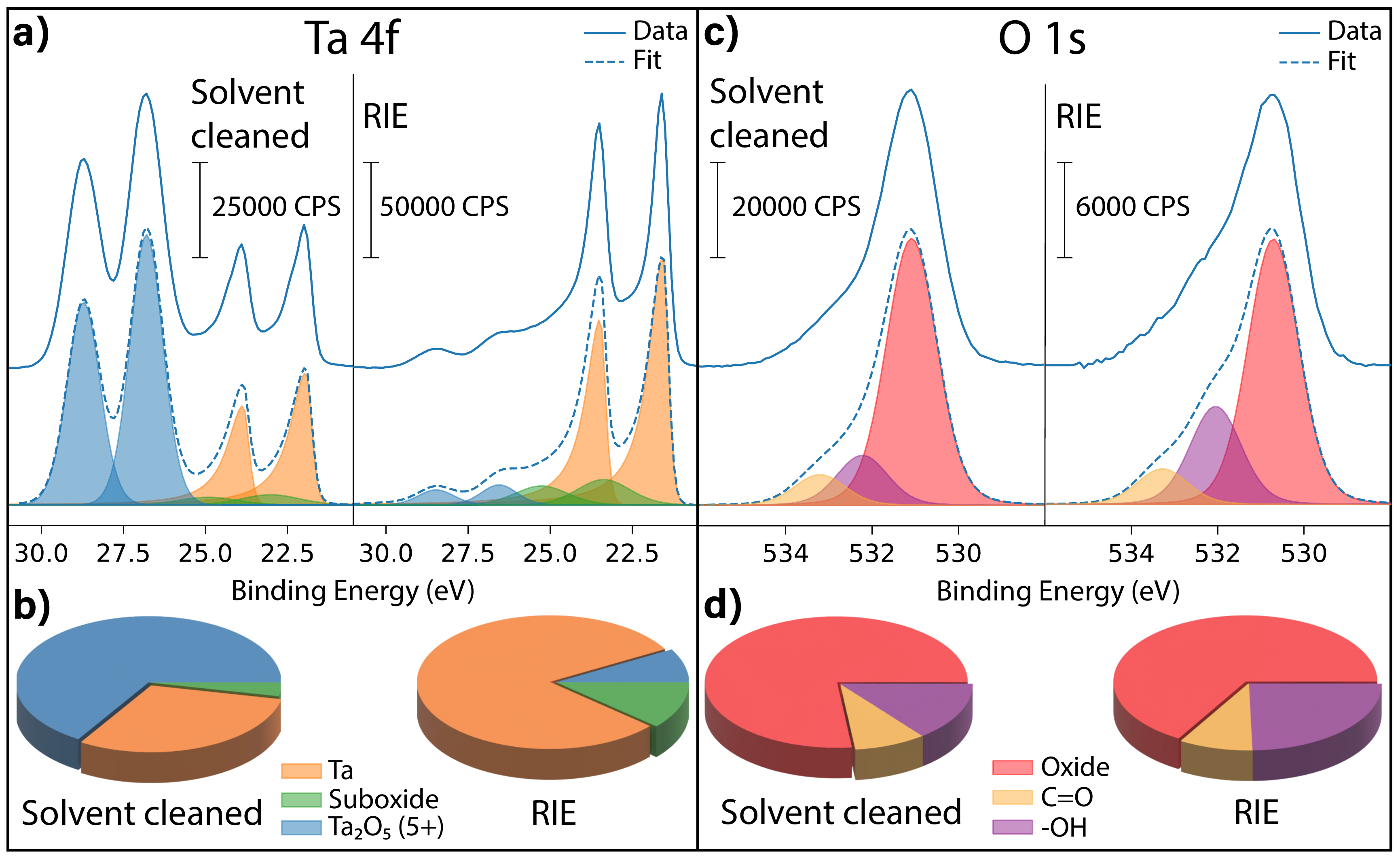}
    \caption{Monochromatic Al K$\alpha$ XPS spectra. Ta 4f (a, b) and O 1s (c, d) levels of the solvent cleaned and RIE processed samples collected at normal emission, and the relative intensities of components used for fitting. Experimental spectra (a, c) are shown as offset solid blue lines, and the fitting envelopes (background subtracted) as dashed blue lines. The Ta 4f core level consists of two dominant sets of spin-orbit split peaks, where the low binding energy doublet (Ta 4f$_{7/2}$, 21.6 eV) corresponds to metallic Ta, while the higher binding energy doublet (Ta 4f$_{7/2}$, 26.7 eV) is Ta$_2$O$_5$. Suboxide components make up a small contribution at intermediate binding energies for most samples. To avoid overfitting, a single symmetric doublet is used to model the suboxide contributions. The O 1s level was fit with contributions from lattice oxygen, surface hydroxyls (-OH) and oxygen-containing organics (C=O).}
    \label{fig4}
\end{figure*}

The surface chemical composition and oxide thickness after etching procedures were tracked by XPS (Figure 4).
Within the probed depth of the solvent-cleaned sample, metallic Ta accounts for $\sim30\%$ of the signal with the remaining intensity due to oxides; a similar trend is observed for the piranha and BOE-treated samples (Figure S1). However, significant changes in the Ta 4f core level are seen after RIE treatments, notably a strong increase in the metallic Ta component. The enhancement of the metallic Ta peak is accompanied by a long shoulder at intermediate binding energies due to suboxides, consistent with previous reports~\cite{wang_situ_2010, simpson_xps_2017, magnuson_electronic_2019}. To gauge the efficacy of these treatments, we estimate the surface oxide thickness after each surface treatment by employing a simple model considering Ta, TaO$_x$, and Ta$_2$O$_5$. From this model, we determine a total oxide thickness of 2.9~nm for the solvent-cleaned sample and 2.8~nm for both the piranha and BOE-treated samples. For the RIE-treated sample, the oxide thickness is significantly reduced to 0.8~nm. 

While the Ta~4f~level has a complex lineshape, the O~1s~level is comparatively simple (Figures 4c,~d). Comparing the solvent cleaned and RIE processed samples, the O~1s retains a similar profile at normal emission, albeit with a reduced relative intensity ($-22\%$). However, by performing grazing angle XPS (Figure S2), we observe a pronounced increase in the surface hydroxyl component for the solvent-cleaned sample. In these more surface sensitive measurements, BOE shows a modest hydroxyl enhancement, while piranha and RIE show no significant increase. The C~1s peak, a general indicator of surface cleanliness, decreases after both piranha ($-22\%$) and RIE ($-8\%$) treatments (Figure S3).

The dramatic change in the Ta~4f core level after RIE processing indicates the removal of surface oxides, a critical step in the fabrication of superconducting devices.
Within a device, the intrinsic losses can be expressed as $F\tan\delta$, where $\tan\delta$ is the loss tangent of the material ($\epsilon''/\epsilon'=\tan\delta$), and $F$ is the filling factor (or participation ratio), defined as the fraction of the total capacitive energy stored in the constituent materials. 
Amorphous oxides have particularly large loss tangents in comparison to their crystalline counterparts.
For example, the loss tangent of amorphous Al$_2$O$_3$ at cryogenic temperatures is $\sim 1.6 \times 10^{-3}$, while crystalline sapphire ($\alpha$-Al$_2$O$_3$) under comparable conditions has a loss tangent of $\sim 2 \times 10^{-8}$~\cite{martinis_decoherence_2005, creedon_high_2011}. While there are currently no literature reports on the low-temperature dielectric loss of crystalline or amorphous TaO$_x$, we expect a similar trend to Al$_2$O$_3$.
The filling factor ($F$) can be engineered to minimize the participation of lossy regions by, for instance, thinning oxide layers at the metal-air interface~\cite{altoe_localization_2022}.
As the metal-air interface is a region subjected to large electric fields, losses due to surface oxides play an outsized role in the final device performance, and any possible reduction in the oxide participation is crucial~\cite{martinis_decoherence_2005}.
%
%
On the solvent-cleaned sample, direct measurement of the boson peak in region B, together with the absence of crystalline oxide phases in GIWAXS, confirms the existence of lossy $a$-TaO$_x$. Furthermore, XPS measurements give evidence for surface-bound hydroxyls, which are likely TLS candidates.
On the RIE-treated sample we observe a reduction in both the oxide layer thickness and the removal of amorphous material, confirming RIE as an effective technique to prepare Ta films for use in superconducting devices.
%
%

In conclusion, our experiments reveal a strong and unexpected near-field optical contrast on Ta films. We attribute these differences to the dissimilar oxidation behavior of 1D nanoridges and flat regions.
%
Probing the two regions with THz nanospectroscopy, we observe an absorption peak across multiple harmonics at 0.5 THz in flat regions, which corresponds to the boson peak, a universal signature of amorphous materials.
This is supported by GIWAXS measurements indicating a lack of crystalline oxide phases on fresh Ta films.
Our observation of the boson peak and its localization to a particular surface region is a rare opportunity to visualize microscopic sources of decoherence.
The boson peak in amorphous materials is directly connected to the TLS invoked in superconducting quantum devices by bistable defects, the major loss channel in qubits.
These defect centers have, to the best of our knowledge, only been observed indirectly via area averaging techniques or in \textit{operando} devices, limiting our microscopic understanding of structure-property relationships.
Our findings further demonstrate that THz s-SNOM can inform and guide the processing of materials for quantum computing~\cite{guo_near-field_2021}.
The direct observation and identification of the boson peak using THz nanospectroscopy opens new doors to explore amorphous materials at the nanoscale.
Finally, we note that these methods are particularly relevant to 2D materials, where reduced dimensionality strongly enhances the near-field optical response, and may be applied to disordered or amorphous forms of these materials such as bilayer SiO$_2$~\cite{freund_controlling_2017}, h-BN~\cite{zhang_structure_2022}, graphene, and the metal dichalcogenides~\cite{ChenC2020, Barnett2021}.
\vspace{10pt}
\section*{Acknowledgements} \label{sec:acknowledgements}

The authors acknowledge that UQ operates on the land of the Jagera and Turrbal peoples. We pay our respects to their Ancestors and their descendants who continue cultural and spiritual connections to Country. The authors acknowledge the facilities, and the scientific and technical assistance, of the Microscopy Australia Facility at the Centre for Microscopy and Microanalysis, The University of Queensland. This work used the Queensland node of the NCRIS-enabled Australian National Fabrication Facility (ANFF). Financial support was provided by the Australian Research Council’s Discovery Projects’ funding scheme (No. DP210103342), the ARC Centre of Excellence for Engineered Quantum Systems (EQUS, No. 286 CE170100009), and the Foundational Questions Institute Fund (Grant No. FQXi-IAF19-04). JS acknowledges support from Maarten B. J. Roeffaers and from the Research Foundation - Flanders (FWO: Grant No. 12Y7221N, V400622N). The authors thank the staff of the BL11 NCD-SWEET beamline at ALBA Synchrotron for their assistance in recording the GIWAXS data.
\section*{Author contributions} \label{sec:authorcontributions}
XG, ZD, BCD, JS, and ES performed the experiments: XG, THz s-SNOM measurements; ZD, XPS and sample etching; BCD, AFM and nano-FTIR measurements; JS, GIWAXS measurements.
XG, ZD, PJ and JS analyzed the data.
ZD, XG and PJ prepared the manuscript, with contributions from all authors.
PJ and ADR supervised the project.
All authors discussed the results and reviewed the manuscript.
\section*{Data availability}
The data that support the findings of this study are available from the corresponding authors upon reasonable request.


\bibliography{references_v3}


\newpage
\setcounter{figure}{0}
\renewcommand{\figurename}{Fig.}
\renewcommand{\thefigure}{S\arabic{figure}}
\section*{Supplementary Information} \label{sec:supp}


\subsubsection*{Surface treatment}
Tantalum on sapphire wafers ($200$ nm Ta, c-plane sapphire; protective photoresist capping layer) were purchased from Star Cryoelectronics. The photoresist coated samples were diced from the same wafer to dimensions of $4 \times 7$ mm$^2$. Photoresist was removed from the diced chips by a 3 min sonication clean in very large scale integration (VLSI) grade acetone, followed by a further 3 min sonication in VLSI IPA. The samples were then rinsed under running DI water ($18.2$ M$\Omega$ cm) for 15 s and blow-dried with nitrogen gas. We refer to samples in this state (photoresist removed, no further treatment) as solvent cleaned. All samples undergo this photoresist removal step before any further surface treatment.

Piranha-cleaned samples were submerged in a 3:1 piranha solution (30 ml H$_2$SO$_4$ to $10$ ml H$_2$O$_2$) for $20$ min, then rinsed in two DI water beakers for 2 min each and dried with nitrogen gas. Buffered oxide etched (BOE) samples were submerged in a buffered HF solution (7:1 ammonium fluoride to hydrofluoric acid) for $150$ s, followed by two separate DI water dips for $15$ s and $10$ s and a nitrogen gas blow-dry.

Reactive ion etching (RIE) was performed in an Oxford Instruments PlasmaPro 80 Reactive Ion Etcher. Etching used carbon tetrafluoride (CF$_4$) gas at a flow rate of $30$ sccm ($20$ mTorr) and a plasma power of $200$ W. An etch rate of $\sim 11$ nm/min was determined by comparing masked and unmasked regions of Ta test samples with a profilometer (Tencor™ P-7 Stylus Profiler). Samples for AFM/SNOM and XPS analysis were etched for 5 min at a plasma power of $200$ W, removing $\sim 55$ nm of material. 

\subsubsection*{X-ray photoelectron spectroscopy}
X-ray photoelectron spectroscopy (XPS) measurements were performed on a Kratos AXIS Supra+ system equipped with monochromated Al K$\alpha$ ($1486.6$ eV) with beam spot size of $300$ $\mu$m. XPS chamber base pressure $<1\times 10^{-8}$ mbar. Survey scans ($0$ - $1200$ eV) and high-resolution scans of the Ta 4f, O 1s, C 1s, and F 1s levels were obtained at emission angles of $0^{\circ}$, $40^{\circ}$ and $70^{\circ}$ for each sample. The Al K$\alpha$ source was used across all samples. Tantalum films were grounded by clips and no sample charging was observed.

XPS analysis was performed in CasaXPS \cite{fairley_systematic_2021}, where peaks were fitted using a symmetrical Gaussian/Lorentzian line-shape and Shirley background subtraction. The Ta 4f peak was fitted with doublet components from metallic Ta and Ta$_2$O$_5$. To use a minimal fitting model that accounts for details seen in angle resolved XPS data across all treatments, we implement an additional doublet at intermediate binding energy to account for suboxide contributions. O 1s spectra were fit with three components corresponding to lattice oxide, surface hydroxyls, and carboxyl groups.

Oxide thickness $d$ was estimated from the intensity ratio of oxide (taken as the sum of Ta$_2$O$_5$ and suboxides) and metal components in the Ta 4f peak by \cite{alexander_quantification_2002}

\begin{equation}
d_{XPS} (\text{nm})=\lambda_o \sin\theta \ln\left(\frac{N_m \lambda_m I_o}{N_o \lambda_o I_m} + 1 \right),
\end{equation}

where $\lambda_m$ ($\lambda_o$) is the inelastic mean free path (IMFP) for electrons in Ta metal (oxide) at a given energy ($\lambda_{\text{Ta2O5}}=2$ nm, $\lambda_m=1.964$ nm), and $\theta$ is the emission angle (measured as the angle between the surface normal and detector). $I_o/I_m$ is the ratio of the oxide ($o$) and metal ($m$) peak intensities measured by XPS at the given emission angle $\theta$, and $N_m/N_o$ is the ratio of volume densities of Ta atoms ($N_m=0.092$, $N_o=0.065$). We limit ourselves to a bilayer model considering a bulk Ta layer, and a mixed oxide layer incorporating both stoichiometric Ta$_2$O$_5$ and suboxides~\cite{li_tuning_2019}.
The intensity of metallic Ta ($I_m$) is determined by the scaled Ta peak area, and the oxide intensity ($I_o$) is taken as the sum of Ta$_2$O$_5$ and suboxide scaled peak areas. 

\subsubsection*{Scattering-type scanning near-field optical microscopy}
In an s-SNOM, a metallic probe tip periodically taps the sample surface, which is simultaneously illuminated by an electromagnetic stimulus, e.g., THz radiation.
The probe tip is transiently polarized by the incident illumination and thereby forms a highly concentrated electric field --- a nanofocus --- near its apex.
As the nanofocus can be positioned with nanometer precision, and the probe tip radius is $\sim60~\textrm{nm}$, THz s-SNOM is able to resolve nanoscale THz responses correlated to AFM sample topography.
Collecting white-light (spectral-averaged) images enables a spatial mapping of the near-field optical response. In contrast, THz nanospectroscopy provides a frequency-dependent imprint of the interrogated material within the nanofocus~\cite{ChenXZ2019Review, Cocker2021Review}.
Therefore, THz s-SNOM is able to bypass the typical Rayleigh diffraction limit and provides nanoscale insight into lattice dynamics and electronic processes, including the characterization of amorphous and crystalline material phases~\cite{ChenC2020, Barnett2021}.

A commercial s-SNOM (neaSNOM, neaspec GmbH, Haar, Germany) is used for near-field experiments with commercial probes (THz: 25PtIr200B-H45, Rocky Mountain Nanotechnology, LLC, Bountiful, USA; nano-FTIR: neaspec nano-FTIR probes). The metallic probe, operating in tapping-mode atomic force microscope, is illuminated by a laser beam for near-field measurements. The tip-scattered field, $E(t)$, was demodulated with high-order harmonics ($n\geq 2$) of the tip oscillation frequency to obtain near-field signals $S_n(t)$.

\textbf{Nano-FTIR:} the beam comes from an integrated laser source (Toptica Photonics AG, Gr{\"a}felfing, Germany) which is generated by a difference frequency mixer where two near-infrared (NIR) pulse trains are superposed. An asymmetric Michelson interferometer is employed to obtain interferograms for mid-IR spectral information. The backscattered field from the tip is directed to a mercury cadmium telluride detector.

\textbf{THz s-SNOM:} a broadband THz radiation ($0$ to $6$ THz) is generated from a photoconductive antenna (low-temperature-grown InGaAs/InAlAs, Menlo Systems GmbH, Martinsried, Germany) under the excitation of a femtosecond NIR ($1560$ nm) laser. The forward tip-scattered component is detected by another photoconductive antenna gated by a NIR pulse identical to the excitation beam. For white-light nanoimaging, the time-delay ($\sim$picosecond) between the excitation and sampling pulse with the strongest THz scattering amplitude is selected for spatially varying spectral-averaged (white-light) THz responses. For THz nanospectroscopy, the point of interest is selected to sweep the optical scanning delay line to obtain the time-dependent THz scattering field for further lock-in detection and Fourier transform.

\textbf{High-aspect-ratio AFM:} a Bruker TESPA-HAR probe was used for tapping-mode AFM imaging.

\subsubsection*{Synchrotron-based grazing incidence wide angle scattering}
Room temperature GIWAXS of Ta films were recorded at NCD-SWEET beamline (ALBA synchrotron in Cerdanyola del Vall\`es, Spain) with a monochromatic ($\lambda = 0.9574$~\AA) X-ray beam of $80 \times 30$~$\mu$m$^2$~[H × V], using a Si (111) channel cut monochromator, which was collimated using an array of Be collimating lenses. The scattered signal was recorded using a LX255-HS area detector (Rayonix, LLC, Evanston, USA) placed at $229.4$~mm from the sample position. Detector tilts, and sample-to-detector distance were calculated using Cr$_2$O$_3$ as calibrant. GIWAXS frames were recorded at incident angles ($\alpha_i$) between $0^{\circ}$ and $0.35^{\circ}$, remaining in the surface-sensitive evanescent regime of scattering and remaining below the full penetration of the Ta film.  Continuous N$_2$ flow over the sample was employed during the measurements to ensure the sample remained in a chemically inert atmosphere during irradiation. Collected 2D images were azimuthally integrated using PyFAI~\cite{ashiotis_fast_2015} and processed using a custom python routine.

\begin{figure*}
    \centering
    \includegraphics[width=\linewidth]{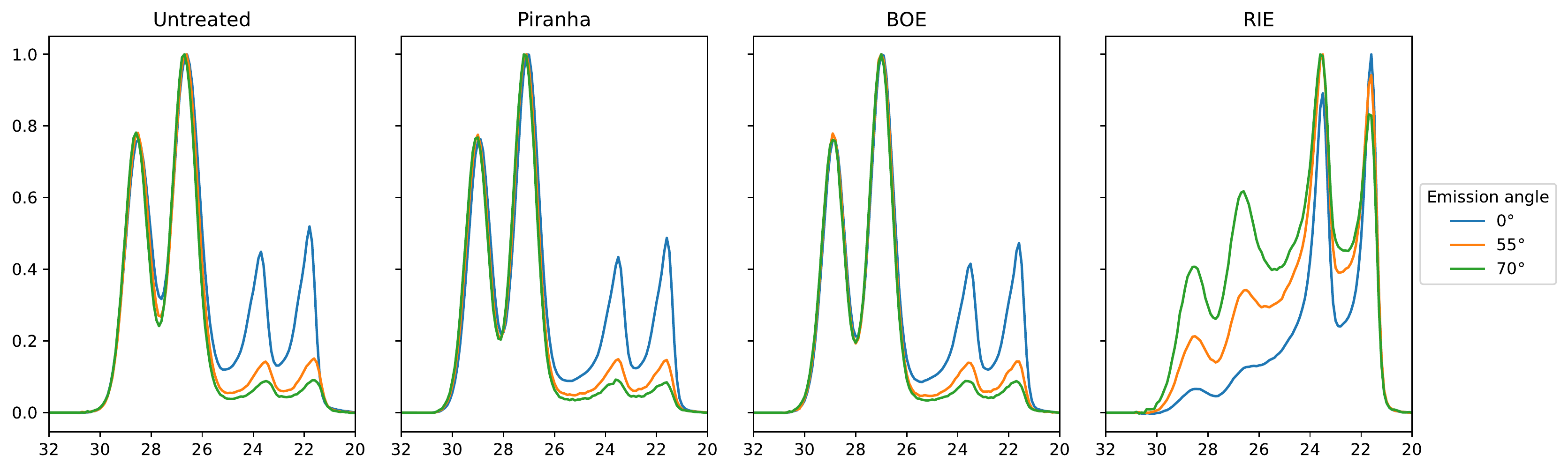}
    \caption{Angle resolved XPS (ARXPS) of the Ta~4f levels.}
    \label{fig:S1}
\end{figure*}

\begin{figure*}
    \centering
    \includegraphics[width=\linewidth]{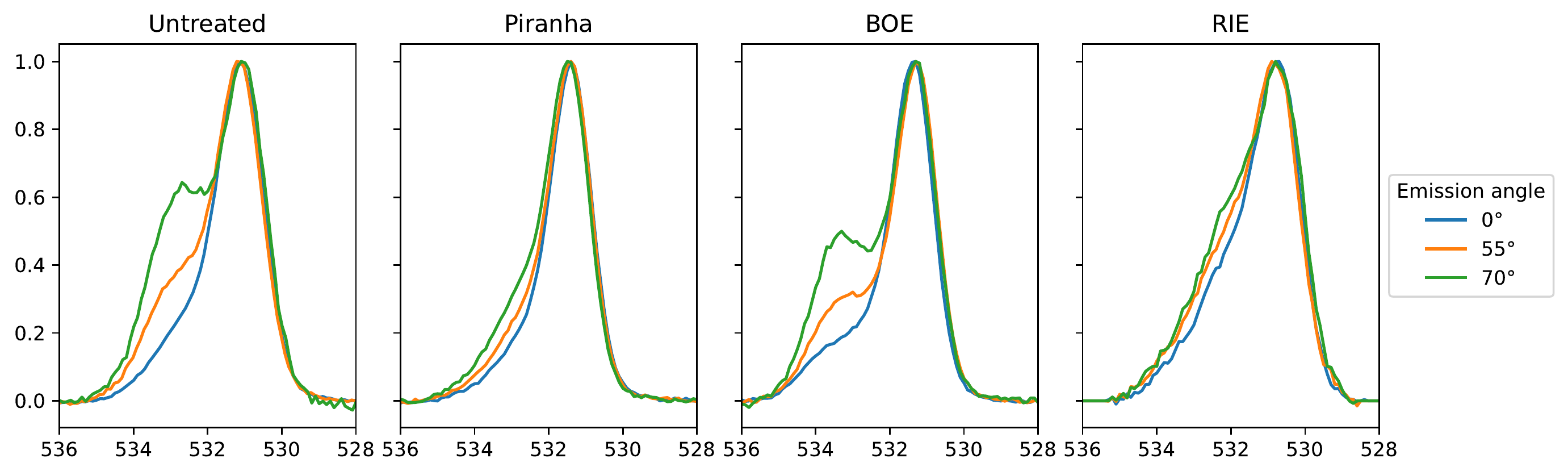}
    \caption{ARXPS of the O~1s levels.}
    \label{fig:S2}
\end{figure*}

\begin{figure*}
    \centering
    \includegraphics[width=\linewidth]{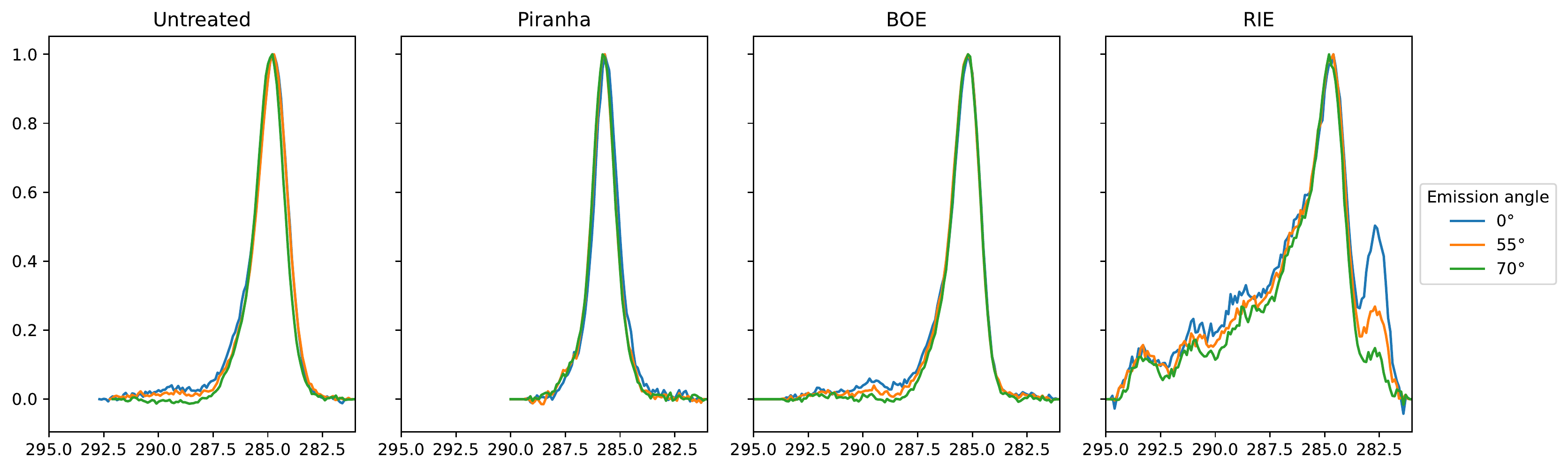}
    \caption{ARXPS of the C~1s levels. On the RIE-etched sample, the C~1s spectrum has components at higher binding energy ($290$ - $295$ eV) which are attributed to C-F bonds formed during etching with CF$_4$.}
    \label{fig:S3}
\end{figure*}

\begin{figure*}
    \centering
    \includegraphics[width=\linewidth]{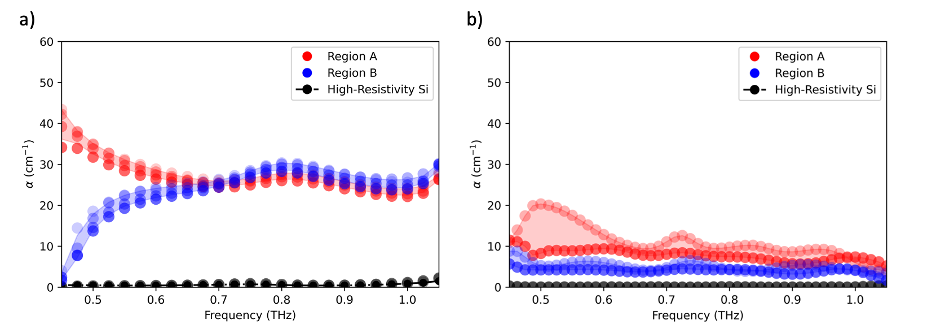}
    \caption{THz spectral absorption. Absorption coefficients without the frequency-dependent normalization on both (a) a solvent-cleaned and (b) an RIE-etched sample on region A (red) and B (blue) with high-resistivity silicon (black) as a reference in the probing THz bandwidths. Near-field scattering signals demodulated at multiple high-order harmonics ($n\geq 2$) of the probe oscillation frequency are used to suppress the background noise. The localized absorption coefficient of the un-etched sample on region B ((a), blue) shares the same characteristic behavior with the reported far-field THz time-domain spectroscopic measurements on a glassy system~\cite{mori_detection_2020}; the same case happens for the normalized absorption coefficient by the square of the frequency (Figure 2b).}
    \label{fig:S4}
\end{figure*}

\begin{figure*}
    \centering
    \includegraphics[width=\linewidth]{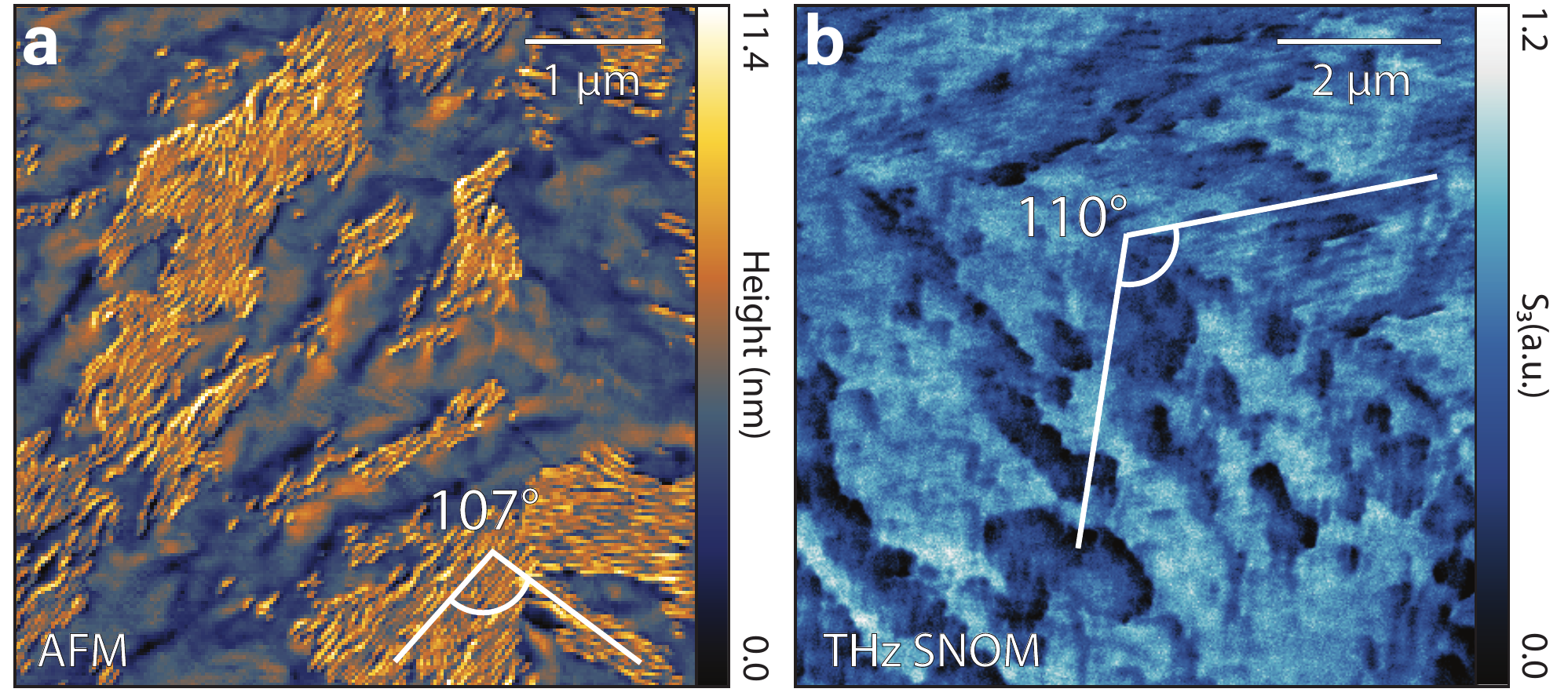}
    \caption{Relative stripe directions at a domain intersection. (a) $5 \times 5$ $\mu$m AFM image and (b) $10 \times 10$ $\mu$m THz nanoimaging ($3^\text{rd}$ harmonic) of nanoridges found in region A meeting at a domain intersection. As expected for bcc Ta, the stripes meet with a relative angle close to $110^{\circ}$. The images were taken with s-SNOM probe tip for THz nanospectroscopy ($\sim60$~nm tip radius).}
    \label{fig:S5}
\end{figure*}

\begin{figure*}
    \centering
    \includegraphics[width=\linewidth]{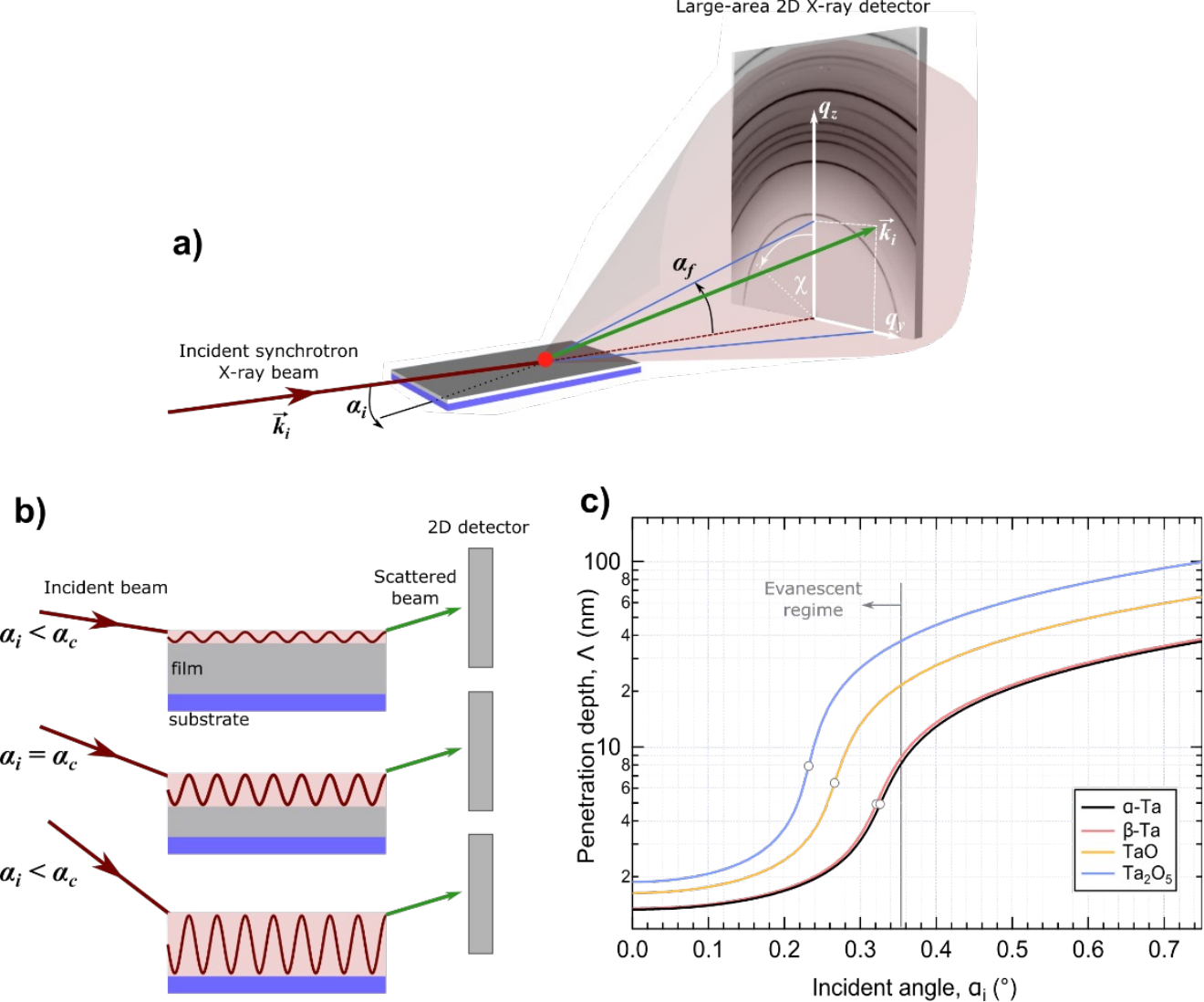}
    \caption{Surface-sensitive study of metallic tantalum and surface tantalum (sub)oxide using synchrotron-based GIWAXS. (a) Schematic illustration of the scattering geometry of synchrotron GIWAXS measurements (ALBA Synchrotron) performed on Ta thin films. The incident X-ray beam ($\lambda = 0.95774$ Å, $12.95$ keV) scatters from the sample under a grazing angle, projecting diffraction signals onto the larger-area imaging detector. Depending on the polycrystalline texture (i.e. orientation and distribution of scattering domains), scattering rings may only appear in a certain azimuthal direction, $\chi$. Integrated data are derived from integrating over the whole image (i.e. $q_{x,y,z}$) are used for the profile analysis in the main text. (b) A schematic illustrating the dependence of penetration depth ($\Lambda$) on the incident angle $\alpha_i$ and critical angle $\alpha_c$. With the Ta and TaO$_x$ materials having a high-frequency refractive index less than 1, the incident X-rays are totally reflected at shallow incident angles ($\alpha_i \leq \alpha_c$), while at higher incident angles they penetrate the material ($\alpha_i > \alpha_c$). The transition between these disparate scattering conditions defines the critical angle ($\alpha_c$), and depends on the X-ray energy and the material dielectric properties. (c) The calculated penetration depth ($12.95$ keV) for Ta (both $\alpha$- and $\beta$-phase material densities~\cite{abadias_elastic_2019}) in comparison to fully (Ta$_2$O$_5$) and partially (TaO) oxidized Ta, which is relatively less dense~\cite{mathieu_aes_1984}. The open circles indicate the calculated critical angle, $\alpha_c$. For the range of angle below and new the critical angle, the X-ray electromagnetic field only interacts a short distance below the film surface due to the evanescent damping ($5$ - $10$ nm), and constitutes the defined evanescent regime. Fortuitously, the less-dense oxide layer acts as a waveguide in this regime, amplifying TaO$_x$-related signals arising from the top surface.}
    \label{fig:S6}
\end{figure*}

\begin{figure*}
    \centering
    \includegraphics[width=0.6\linewidth]{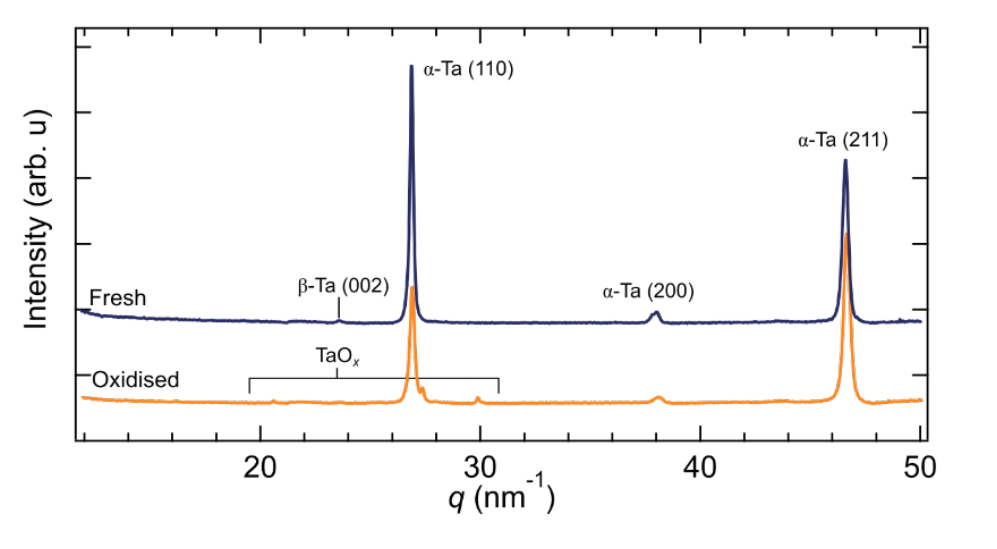}
    \caption{Full 1D scattering profile of integrated GIWAXS signals recorded from Ta films exhibiting a fresh (solvent-cleaned) and partially oxidized (aged) surface. }
    \label{fig:S7}
\end{figure*}

\begin{figure*}
    \centering
    \includegraphics[width=\linewidth]{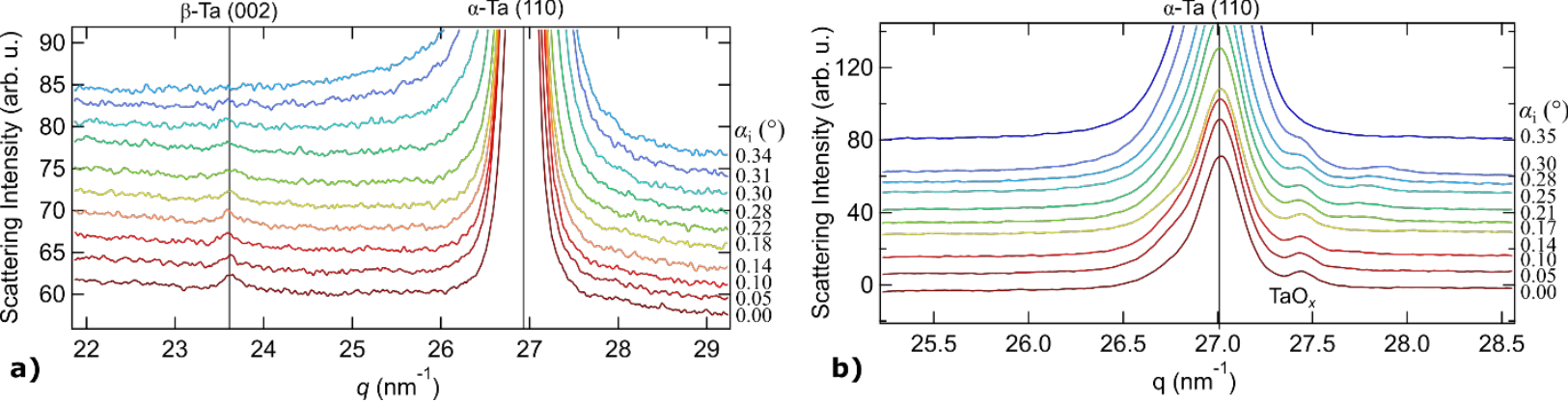}
    \caption{Surface depth profiles of Ta films. Integrated GIWAXS profiles recorded as a function of incident angle (values inset) from (a) fresh and (b) an aged solvent-cleaned Ta film. As the angle of incidence ($\alpha_i$) nears the critical angle of the Ta metal film ($\sim 0.32^{\circ}$), the $\beta$-Ta phase signals disappear in a), indicating they arise from the surface rather than the bulk. Likewise, while $\alpha_i<0.34$ the relatively less dense TaO$_x$ has its scattering signal amplified from waveguide-like effects, before the angle is high enough to stop reflecting off the Ta surface and penetrate the bulk.}
    \label{fig:S8}
\end{figure*}

\begin{figure*}
    \centering
    \includegraphics[width=0.8\linewidth]{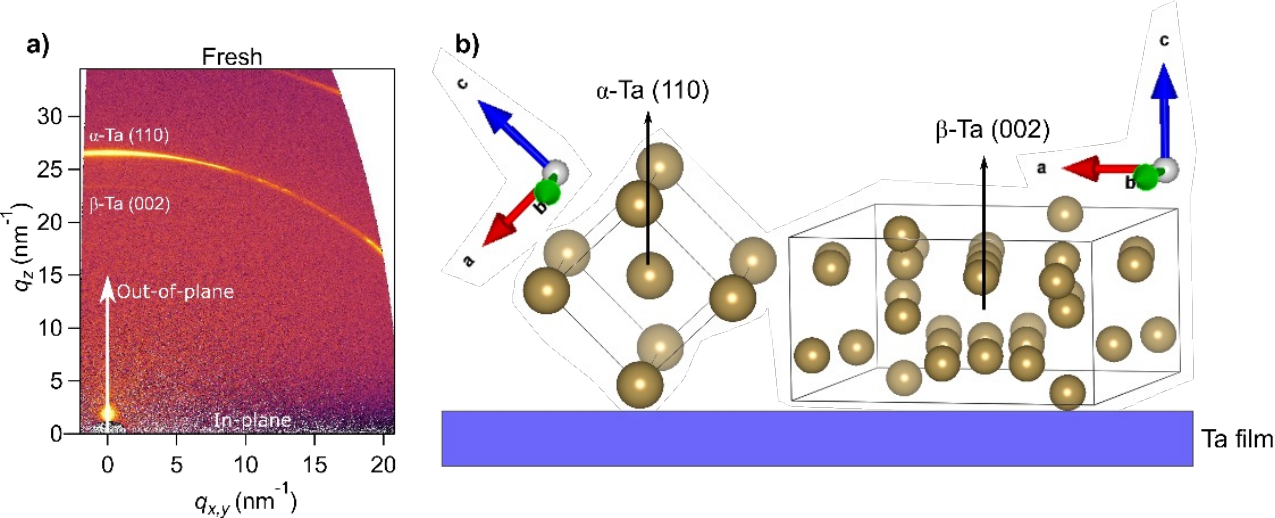}
    \caption{Orientation and distribution of $\beta$-Ta and $\alpha$-Ta crystal in the film. (a) Relative orientation of $\alpha$-Ta (110) and $\beta$-Ta (002) Bragg scattering planes incident on the 2D $q_z$ vs $q_{x,y}$ image, i.e., the so-called “missing wedge” is absent in these data for simplicity. (b) Geometric illustrations of the relative crystal texture on the substrate, showing the most probable orientations of the measured $\beta$-Ta and $\alpha$-Ta crystal structure.}
    \label{fig:S9}
\end{figure*}

\end{document}